%% file: main.tex
\documentclass[10pt,conference]{IEEEtran}
\IEEEoverridecommandlockouts
\usepackage{cite}
\usepackage{amsmath,amssymb,amsfonts}
\usepackage{algorithmic}
\usepackage{graphicx}
\usepackage{textcomp}
\usepackage{url}
\usepackage{xcolor}
\usepackage[]{collab}
\usepackage{makecell}
\usepackage{enumitem}
\usepackage{booktabs}
\usepackage{multirow}
\usepackage{listings}
\usepackage{balance}
\usepackage{xcolor}
\usepackage{tabularx}

\def\BibTeX{{\rm B\kern-.05em{\sc i\kern-.025em b}\kern-.08em
    T\kern-.1667em\lower.7ex\hbox{E}\kern-.125emX}}

\newcommand\name{\textsc{MicroRacer}\xspace}

\collabAuthor{zb}{teal}{Zhuangbin Chen}
\collabAuthor{zl}{blue}{Zhiling Deng}
\collabAuthor{jp}{cyan}{Juepeng Wang}

\begin{document}

\title{\name: Detecting Concurrency Bugs for Cloud Service Systems}

\author{
\IEEEauthorblockN{Zhiling Deng}
\IEEEauthorblockA{
Sun Yat-sen University\\
Zhuhai, China \\
dengzhling3@mail2.sysu.edu.cn}
\and
\IEEEauthorblockN{Juepeng Wang}
\IEEEauthorblockA{
Sun Yat-sen University\\
Zhuhai, China \\
wangjp39@mail2.sysu.edu.cn}
\and
\IEEEauthorblockN{Zhuangbin Chen$^{\dagger}\thanks{$^{\dag}$Zhuangbin Chen is the corresponding author.}$}
\IEEEauthorblockA{
Sun Yat-sen University\\
Zhuhai, China \\
chenzhb36@mail.sysu.edu.cn}
}

\maketitle

\begin{abstract}

Modern cloud applications delivering global services are often built on distributed systems with a microservice architecture.
In such systems, end-to-end user requests traverse multiple different services and machines, exhibiting intricate interactions.
Consequently, cloud service systems are vulnerable to \textit{concurrency bugs}, which pose significant challenges to their reliability.
Existing methods for concurrency bug detection often fall short due to their intrusive nature and inability to handle the architectural complexities of microservices.
To address these limitations, we propose \name, a non-intrusive and automated framework for detecting concurrency bugs in such environments. 
By dynamically instrumenting widely-used libraries at runtime, \name collects detailed trace data without modifying the application code.
Such data are utilized to analyze the happened-before relationship and resource access patterns of common operations within service systems.
Based on this information, \name identifies suspicious concurrent operations and employs a three-stage validation process to test and confirm concurrency bugs.
Experiments on open-source microservice benchmarks with replicated industrial bugs demonstrate \name's effectiveness and efficiency in accurately detecting and pinpointing concurrency issues.

\end{abstract}


\input{sections/01-intro}
\input{sections/02-background}
\input{sections/03-methodology}
\input{sections/04-implementation}
\input{sections/05-experiment}
\input{sections/06-related_work}
\input{sections/07-conclusion}

\section*{Acknowledgement}

This work is supported by the National Natural Science Foundation of China (No. 62402536).

\balance
\bibliographystyle{IEEEtran}
\bibliography{bibliography}


\end{document}

%% file: sections/01-intro.tex
\section{Introduction}

Cloud service systems, characterized by their distributed nature and microservice architecture, offer unparalleled scalability and flexibility.
This architectural approach empowers businesses to innovate rapidly and deliver superior user experiences.
For instance, Uber operates with several thousand microservices~\cite{DBLP:conf/sigsoft/HeFLZ0LR023}, enabling it to efficiently provide seamless services across different regions.
Similarly, WeChat hosts over 3,000 microservices to handle billions of daily requests worldwide~\cite{DBLP:conf/cloud/ZhouCLWSLGOY18}.
With this level of complexity and scale, ensuring the reliability of cloud service systems becomes paramount~\cite{DBLP:conf/sigsoft/ChenKLZZXZYSXDG20}.

In such systems, the end-to-end request flows are inherently complex, involving interactions among multi-tier services and machines.
Due to the large number of simultaneous execution of multiple operations, cloud service systems often suffer from \textit{concurrency bugs} \cite{DBLP:conf/eurosys/TangBZKJGX23,DBLP:conf/sigsoft/ChenKLZZXZYSXDG20}, which could lead to inconsistent states, data corruption, and ultimately system failures.
Modern cloud services typically employ a loosely coupled design, with each maintaining its own independent datastore~\cite{DBLP:conf/sosp/LoffP0M023}.
A single user request can make multiple read and write operations across different datastores (owned by different services) during the course of its execution.
This further compounds the complexity to managing data consistency and system reliability.

Concurrency bug detection is a well-studied topic in distributed systems~\cite{DBLP:conf/asplos/Leesatapornwongsa16, DBLP:journals/tos/LiZLGGHL23} and web applications~\cite{DBLP:conf/eurosys/LiCLLZGGLL18, DBLP:conf/asplos/LiuWLLYT18, DBLP:conf/sosp/LoffP0M023}.
However, traditional distributed systems, such as Hadoop and Zookeeper, typically do not involve independent datastores for each component.
Meanwhile, race detection in web applications addresses the issue from the perspective of external client requests, often neglecting the internal complexities and architectural characteristics of cloud service systems.
As a result, existing methods fail to consider these unique architectural aspects, rendering them less effective in the context of microservice environments.
Moreover, these tools are often intrusive~\cite{DBLP:conf/sac/KochSJP20, DBLP:conf/sigsoft/QiuS0J21}.
They instrument application source code to monitor and control execution, which can introduce significant overhead and potentially alter the behavior of the system being analyzed.
Therefore, there is a pressing need for non-intrusive and effective techniques tailored to the specific challenges posed by microservice architectures in cloud service systems.
We introduce the primary challenges as follows:

\begin{itemize}[noitemsep,leftmargin=5.5mm]

    \item \textbf{Complex Communication Patterns}:
    Microservice systems involve a variety of communication patterns, such as request-response, event-driven, and message queues.
    This introduces non-deterministic execution orders and complicated timing relationships that are difficult to reason about. Standard happen-before models~\cite{DBLP:conf/asplos/LiuLLLLGT17,DBLP:conf/asplos/LiuWLLYT18} may not be sufficient to capture these complex, application-level communication semantics, making it challenging to accurately determine which operations are truly concurrent.

    \item \textbf{Decentralized State and Cross-service Inconsistency}: Microservices promote a ``database-per-service'' design, resulting in multiple independent and ``mutually-oblivious'' datastores~\cite{DBLP:conf/sosp/LoffP0M023}. A single end-user request often runs across several services, performing writes to these different datastores. This creates an implicit causal dependency across services that traditional consistency models, which operate within a single datastore, cannot see or enforce.

    \item \textbf{Accuracy and Automation in Bug Validation}: An effective concurrency bug detection method must be able to achieve high accuracy in identifying concurrency issues, minimizing both false positives and false negatives. Moreover, the bug validation process should be automated to reduce the need for extensive human intervention, such as writing assertions manually to check the correctness of system behaviors~\cite{DBLP:conf/icdcs/HeorhiadiRJRS16,DBLP:conf/cloud/MeiklejohnESMP21,DBLP:conf/icws/LongWCCCW20}.

\end{itemize}

To address these challenges, we propose \name in this work, a concurrency bug detection framework for microservice systems, which is non-intrusive and automated.
At a high level, \name achieves this goal by first identifying suspicious pairs of concurrent operations based on trace analysis, and subsequently testing for concurrency issues by reversing their execution order. 
If reversing the order leads to different results or errors, it indicates the presence of potential concurrency bugs.

\name tackles the first challenge by employing dynamic instrumentation to automatically instrument widely-used libraries at runtime, rather than modifying the application's code.
We focus on the libraries for common operations in cloud services, such as request/response handling, database interactions, message queuing, among others.
This approach facilitates the collection of trace data,
which capture the happened-before relationship of these operations and their resource access patterns.
Based on such information, \name analyzes concurrent behaviors between different microservices and identifies potential issues.
For the second challenge, \name carefully examines the resource access patterns among different request flows.
Only pairs that access the same resource entities with at least one write operation will be selected.
We also employ pruning strategies to narrow down the search space by considering both shared state locality and the causal dependencies in a request flow.
To address the last challenge, \name executes a three-stage validation process to automatically ascertain whether a pair of concurrent operations triggers a bug.
Particularly, to accelerate this process, \name performs an early validation to preliminarily uncover pairs with obvious conflicts.

We demonstrate the practical effectiveness of \name through experiments with 14 industrial concurrency bugs replicated on four open-source microservice benchmarks, 
i.e., TrainTicket~\cite{DBLP:journals/tse/ZhouPXSJLD21}, Bank of Anthos~\cite{Bank-of-Anthos}, and two applications from DeathStarBench~\cite{DBLP:conf/asplos/GanZCSRKBHRJHPH19}.
Our evaluation shows that \name is able to accurately identify concurrent operations in microservices and effectively pinpoint bugs.

To sum up, we make the following major contributions:

\begin{itemize}
    \item We propose \name, a non-intrusive framework for automated concurrency bug detection in microservice systems.
    \name instruments widely-used libraries to monitor the happened-before relationship among common service operations and their resource access patterns.
    It then locates the suspicious concurrent operations and validates their correctness by controlling their execution order.

    \item We conduct experiments with popular open-source microservices benchmarks, where we replicate industrial bug cases.
    The experimental results demonstrate the efficiency and effectiveness of \name in revealing concurrency bugs in microservices systems.
    The implementation of \name and bug replications are publicly available\footnote{\url{https://github.com/OpsPAI/MicroRacer}}.
\end{itemize}

%% file: sections/02-background.tex
\section{Background}
\label{sec:background_motivation}

\subsection{Concurrency Issues in Cloud Services}
\label{sec:background_concurrency_issues}

Cloud service systems predominantly adopt architectural patterns that emphasize the loose coupling of services.
Each service can handle requests independently, enabling the system to distribute the load and process multiple requests in parallel.
In such systems, the execution flows of user requests often involve complex interactions among multiple services.
As a result, concurrency issues can arise, posing significant threats to system reliability and performance.

Compared to traditional distributed systems and web applications, concurrency issues in cloud service systems are exacerbated by several factors.
First, the microservices architecture, which promotes independent databases and network-based communication among services, introduces challenges such as inconsistent states, partial failures, and intricate dependency chains.
The fact that cloud services may span multiple data centers or even regions further complicates matters with network latency, partitioning, and synchronization issues.
Moreover, many cloud services adopt eventual consistency models to enhance performance and scalability, leading to scenarios where different parts of the system hold divergent views of the data.
This inconsistency can cause concurrency issues when services interact.
Asynchronous communication mechanisms, such as message queues and event streams, introduce non-deterministic execution orders and timing challenges, adding another layer of complexity.

Therefore, to detect concurrency bugs in such systems, we must be able to effectively monitor the key points of interaction between services to log relevant states and events, such as database accesses, message queue operations, and inter-service communications.
These data serve as the basis for identifying patterns and anomalies that indicate potential concurrency issues.

\subsection{Distributed Tracing}

As discussed in Sec.~\ref{sec:background_concurrency_issues}, the complexity of inter-service communication and the dynamic nature of microservices architecture require robust mechanisms for observability and monitoring.
One such mechanism is distributed tracing, which plays a pivotal role in understanding the behavior and performance of distributed applications~\cite{DBLP:conf/nsdi/FonsecaPKSS07,sigelman2010dapper,DBLP:conf/sosp/MaceRF15,DBLP:conf/sosp/KaldorMBGKOOSSV17,DBLP:conf/sigcomm/ShenZXSLSZWY0XL23}.
A trace is a structured record that captures the hierarchical sequence of events that occur in multiple nodes or components, forming a tree-like representation of the request's journey through the system~\cite{10.1145/3728888,DBLP:conf/nsdi/ZhangXAVM23}.
It is composed of multiple spans, which are the basic units of work that represent a single, indivisible operation within a service.
Each span includes critical information such as timestamps, service logs, and identifiers that link it to other spans within the same trace.

The concept of tracing has been extensively employed to understand system behavior, particularly for reliability analysis~\cite{DBLP:conf/nsdi/WuCP19} and performance bottleneck identification~\cite{DBLP:conf/cloud/LuoXLYXZDH021,DBLP:conf/usenix/0002RRPSC22}.
For instance,~\cite{DBLP:conf/asplos/LiuLLLLGT17,DBLP:conf/asplos/LiuWLLYT18,DBLP:conf/eurosys/LiCLLZGGLL18,DBLP:conf/sigsoft/QiuS0J21} have leveraged distributed tracing to detect and resolve concurrency issues, based on static and dynamic analysis techniques.
In these works, static analysis involves injecting tracing logic into the application code to collect state data, which helps identify potential concurrency issues by examining the code structure and execution paths;
while dynamic analysis requires specifying operations such as thread-related activities, RPC interfaces, and socket communication for tracing, allowing for real-time monitoring and capturing of the system's behavior under various load conditions.
Therefore, they often necessitate manual instrumentation, which is intrusive and ad-hoc, thereby limiting the generalizability and completeness of the collected data.

In this work, we explore a unified approach of distributed tracing that is non-intrusive to the application.
By employing bytecode-level instrumentation, we intercept calls to common libraries used by applications and inject tracing code dynamically.
The traces are then used to construct requests flows that reflect the architecture of microservices.
This method allows for the collection of traces without modifying the application code and enhances the detection of concurrency bugs.

\begin{figure}
    \centering
    \includegraphics[width=0.87\linewidth]{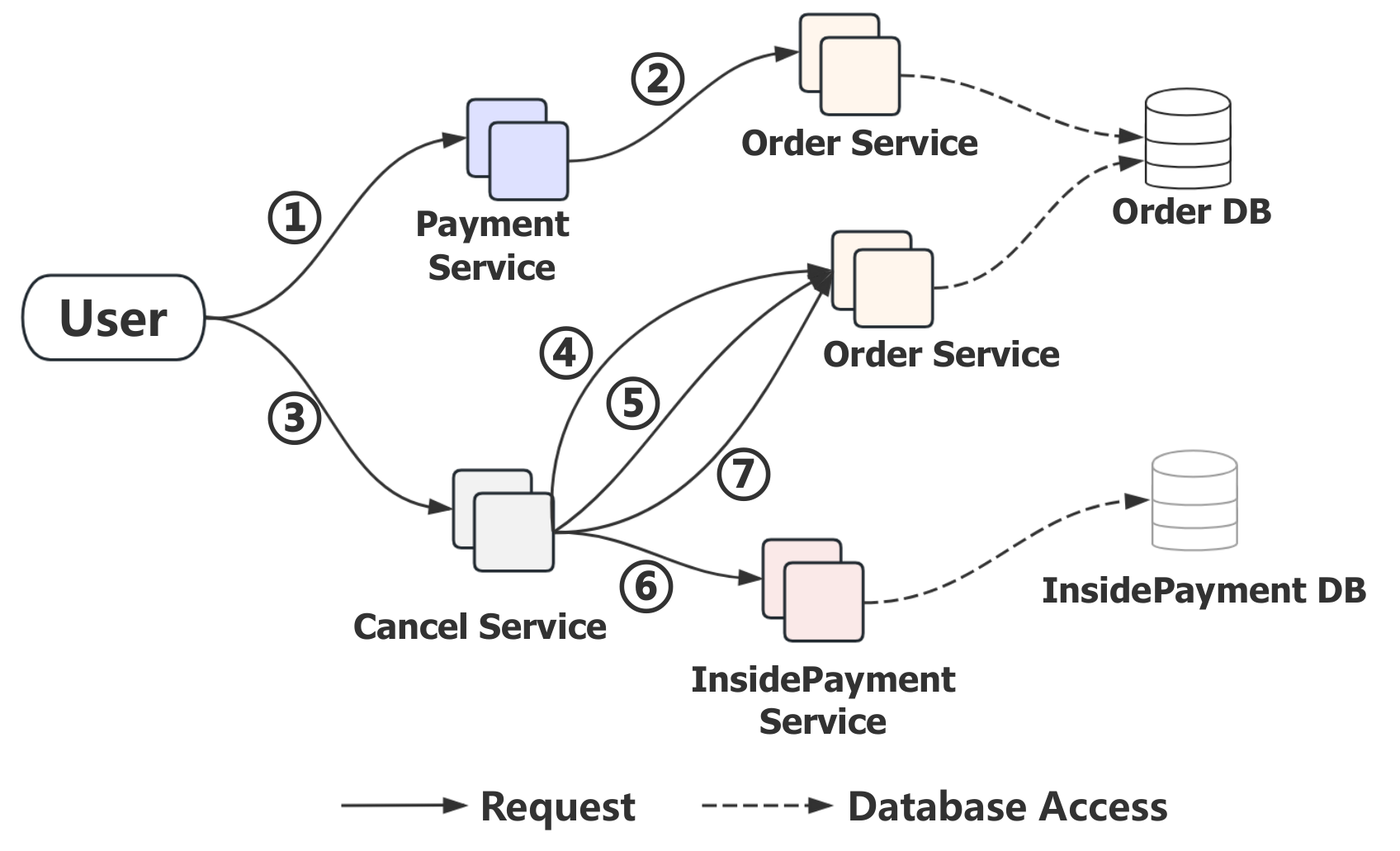}
    \vspace{-5pt}
    \caption{Service call graph of the illustrating example}
    \label{fig:Illustrating Example}
    \vspace{-4pt}
\end{figure}

\begin{figure}
    \centering
    \includegraphics[width=1\linewidth]{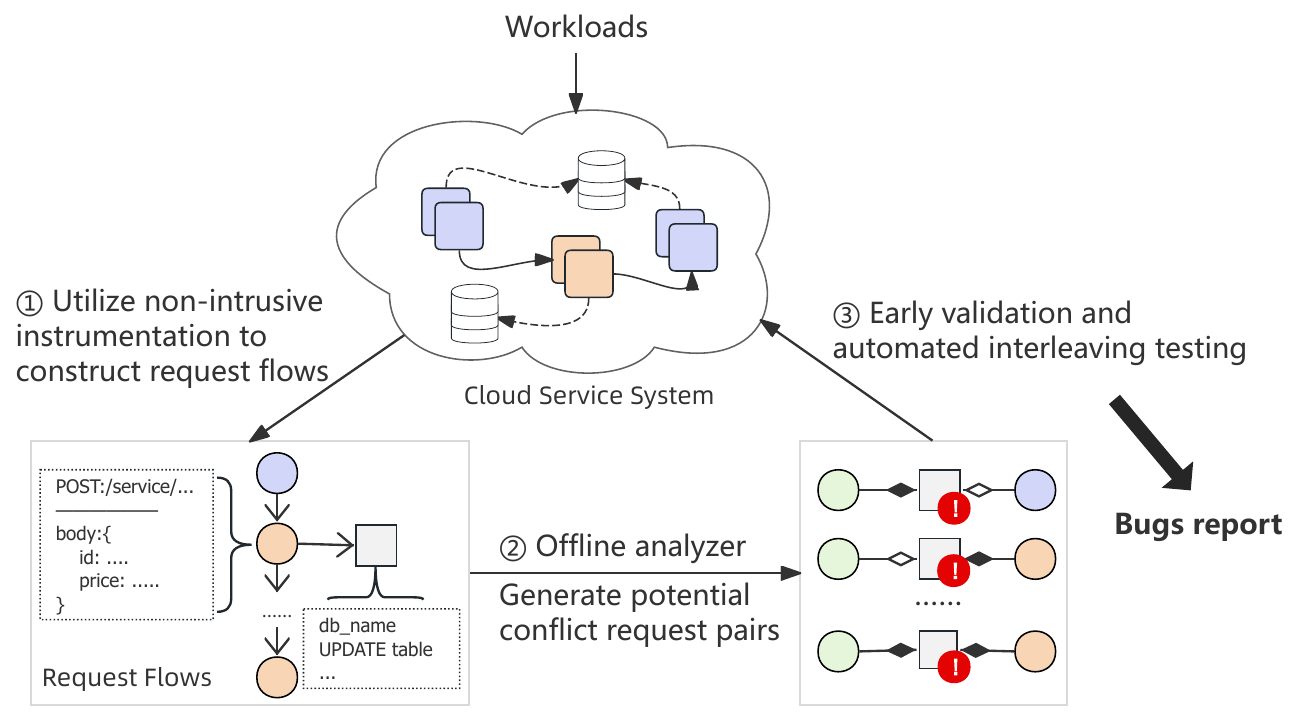}
    \vspace{-6pt}
    \caption{The architecture of \name}
    \label{fig:Framework}
    \vspace{-6pt}
\end{figure}



\subsection{An Illustrating Example}
\label{Illustrating Example}

In this section, we present an example of concurrency bug, which is a functional fault reported by an industrial survey~\cite{DBLP:journals/tse/ZhouPXSJLD21}.
The bug has been replicated in the TrainTicket~\cite{TrainTicket:F13}, a widely-used benchmark in microservice-related research.



Fig.~\ref{fig:Illustrating Example} depicts the service call graph for this bug, which originates from a race condition between two user-initiated operations, i.e., a payment and a subsequent cancellation for the same order. From the perspective of our framework, these actions trigger two distinct, concurrently executing request flows, i.e., flow A (payment): \textcircled{1}→\textcircled{2} and
flow B (cancellation): \textcircled{3}→\textcircled{4}→\textcircled{5}→\textcircled{6}→\textcircled{7}.
The core flaw lies in the system's optimistic assumption that the payment flow (A) will always complete before the cancellation flow (B) begins. However, in a distributed environment, factors like network latency and service scheduling make execution order non-deterministic. This creates a race where both flows compete to read and write to the same order entity in the Order DB.
Let us trace the intended and the buggy execution flows:

\begin{itemize}[noitemsep,leftmargin=5.5mm]
    \item \textbf{Initial State:} The order exists in the Order DB with a status of 0 (\texttt{Not Paid}).

    \item \textbf{Intended (Sequential) Execution:} The Payment Service would process the request, calling the Order Service (\textcircled{2}) to update the order status to \texttt{Paid}. Subsequently, the Cancel Service would handle request \textcircled{3}, read the \texttt{Paid} status, and correctly initiate a refund.

    \item \textbf{Buggy (Concurrent) Execution:} Due to non-deterministic factors like network latency, the cancellation flow begins before the payment flow completes. The Cancel Service invokes the Order Service (\textcircled{4}), which reads the order's state. Because the payment is still pending, it retrieves the stale status \texttt{Not Paid}. Based on this outdated information, the Cancel Service's business logic incorrectly concludes that no refund is required. It then proceeds with the cancellation, updating the order status first to 7 (\texttt{Refunding}) via request \textcircled{5} and finally to 4 (\texttt{Cancelled}) via request \textcircled{7}. Concurrently, the Payment Service eventually processes request \textcircled{2}, successfully charging the user and updating the order in the database.
\end{itemize}

The result is a critical data consistency failure: the user is charged for the order, but the system fails to issue a refund because the cancellation logic operated on stale data. The fault appears to be a classic server-side request race, but its root cause is deeper, originating from the distributed nature of the architecture. The conflict is an inter-service race condition, where multiple services (Payment Service, Cancel Service) concurrently access and modify a shared resource (the order entity) managed by another service (Order Service). This scenario highlights the difficulty of maintaining correctness when state transitions are coordinated across independent, ``mutually-oblivious'' services, a primary challenge \name is designed to detect.

%% file: sections/03-methodology.tex
\section{Methodology}
\label{sec:methodology}
\subsection{Overview}

To detect concurrency bugs in service systems similar to those implied in Fig.~\ref{fig:Illustrating Example}, we propose \textsc{MicroRacer}, 
a framework that tracks request flows in cloud service systems based on non-intrusive instrumentation and automates the detection of request concurrency bugs. 
The architecture of \name is shown in 
Fig.~\ref{fig:Framework}, which consists of three phases, i.e.,
\textit{request flow tracing}, 
\textit{conflict request pair identification}, 
and \textit{automated bug validation}.

\begin{figure}
    \centering
    \includegraphics[width=0.95\linewidth]{figures/RequestFlow.pdf}
    \caption{Request flows of the illustrating example. The flow A (1→2) and the flow B (3→4→5→6→7) have race conditions at data spans \texttt{b} and \texttt{d}.}
    \label{fig:RequestFlow}
\end{figure}

The first phase uses non-intrusive instrumentation to capture end-to-end request flows as the system executes workloads. This process records detailed traces of inter-service calls and operations on shared state resources (e.g., databases, caches). The second phase processes these traces to link requests to the specific data entities they access. It then identifies potential conflicts, i.e., pairs of requests that access the same entity where at least one is a write operation. To improve efficiency, we apply static pruning strategies based on state store locality and request flow causality to significantly narrow the search space. The final phase confirms true positives. It first performs a fast-pass early validation for obvious bugs. The remaining suspicious pairs then undergo automated interleaving testing, where the requests are replayed in both original and reversed orders. A bug is confirmed if this reordering produces a different outcome, as verified by a testing oracle.

\subsection{End-to-end Request Flow Construction}
\label{sec:End-to-end Request Flows Construction}


To detect concurrency bugs, \name must understand the interactions between services and their access to shared states. This is achieved by recording inter-service calls (e.g., via REST APIs, RPC, or message queues) and operations on shared state resources (e.g., databases, caches, object stores) at runtime. From these data, it builds end-to-end request flows that link the original user requests to the specific stateful operations they trigger.
These detailed traces are fundamental for identifying potential conflicting request pairs and corresponding test cases, which are crucial for helping developers find the root cause of the bugs reported by \name.

\name records two types of runtime information, i.e., \textit{data span} and \textit{request span}. When the system receives a user request, \name assigns it a unique flow ID. This flow ID is propagated across service calls, linking individual request spans into a coherent request flow. A data span represents a single, atomic operation on a shared state resource. Each request span can be associated with one or more data spans, depending on how many times the service accesses shared states. Consequently, a request flow starts with a user request and chains together the sequence of inter-service calls and shared state interactions, as illustrated in Fig.~\ref{fig:RequestFlow}.

\subsubsection{Instrumentation for Data Span}

To record operations on shared state, \name instruments different client libraries and SDKs used to interact with various state stores. This approach captures a complete picture of the potential conflicts in modern microservice systems. A data span is created for each interaction, tagged with the propagated flow context to link it back to the originating request.

\begin{itemize}[noitemsep,leftmargin=5.5mm]
    \item \textit{Databases}: For traditional databases, we instrument commonly-used interfaces like JDBC. \name records the database connection information and the full SQL statement, which are used to infer the specific database entities being accessed (e.g., by parsing the primary keys from \texttt{WHERE} clauses).

    \item \textit{Message Queues}: For message queue systems such as Kafka and RabbitMQ, \name instruments the producer and consumer client libraries. It captures the topic/queue name, partition key, and relevant message headers or payload identifiers. A \texttt{produce} operation is treated as a write to the shared state (the topic), while a \texttt{consume} operation is treated as a read.

    \item \textit{Caches}: For distributed caches like Redis or Memcached, the instrumentation targets the client APIs. \name records the key being accessed and the type of operation (e.g., \texttt{GET} for a read; \texttt{SET}, \texttt{DEL} for a write).

    \item \textit{Shared Storage}: For object stores like Amazon S3 or shared file systems, \name instruments the corresponding SDKs. It records the bucket and object key (or file path) as the resource identifier and classifies operations like \texttt{getObject} as reads and \texttt{putObject} as writes.
\end{itemize}

By instrumenting these various client libraries, \name creates data spans that associate each state access operation with the correct request flow.

\subsubsection{Instrumentation for Request Span}

To trace inter-service communication, \name automatically instruments the library APIs used for various protocols, in contrast to previous studies~\cite{DBLP:conf/sigcomm/LiCL19, DBLP:conf/sose/CernyABMT22a, DBLP:conf/closer/BushongDC22} based on static analysis.
While this section focuses on REST APIs, the same principles apply to other communication protocols like gRPC or asynchronous messaging, which can be traced by instrumenting their respective library APIs.
Specifically, when a service makes an REST API call to another, it typically uses standard HTTP client libraries. \name instruments these libraries to record request information (headers, body) before and after the call, forming a request span. To enable cross-service tracing, \name injects the flow context (containing the flow ID) into the request headers, following standard distributed tracing practices. It also instruments server-side request handlers to receive and maintain this context.

\begin{figure}
    \centering
    \includegraphics[width=0.95\linewidth]{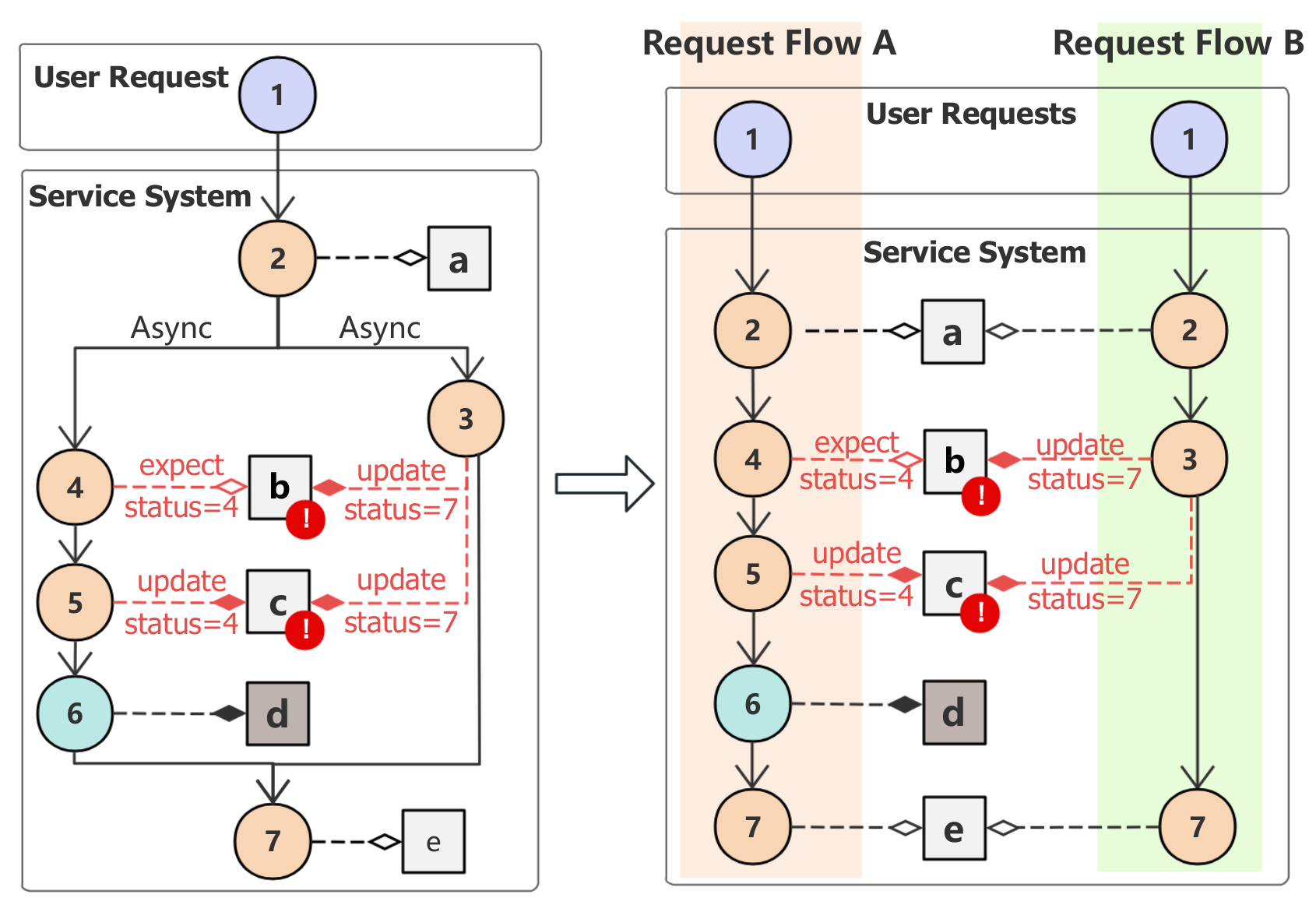}
    \caption{Request flows when a service asynchronously invokes other services. The left diagram shows the actual request flow in the system, while the right diagram shows the request flow split by \name during analysis.}
    \label{fig:AsyncFlow}
\end{figure}

In asynchronous scenarios where a service spawns new threads for parallel processing, \name instruments common asynchronous frameworks (e.g., Spring Async Annotation). It wraps asynchronous tasks (like \texttt{Callable} objects) with the flow context, ensuring that requests initiated from these new threads are correctly associated with the original request flow.
A request may diverge when a main thread and sub-threads initiate requests concurrently, and later converge when the main thread waits for the sub-threads to complete. To simplify conflict analysis, we split such complex flows into multiple, sequential request flows. 
For example, in Fig.~\ref{fig:AsyncFlow}, the service handling request \textcircled{2} starts a sub-thread that sends request \textcircled{3} in parallel with the main thread's requests \textcircled{4}, \textcircled{5}, and \textcircled{6}, then waits for the sub-thread to complete before sending request \textcircled{7}. \name regards this as two distinct request flows (as shown on the right), allowing it to correctly identify potential conflicts between requests like \textcircled{3} and \textcircled{4}.

Besides service invocations and intra-request parallelism, \name also identifies inter-request synchronization through library instrumentation of distributed locks and message queues. 
For distributed locks, it records the lock identifier and operations (such as \texttt{SETNX} in Redis) to determine the mutual exclusion relationship between request flows. For message queues, it links the request flow that sends a message to the request flow that processes it, treating the message passing as a synchronization point. 
This comprehensive instrumentation enables \name to accurately model a wide range of common, explicit synchronization patterns, i.e., service calls, internal parallelism, distributed locks, and message queues.

\subsubsection{An Example}

Fig.~\ref{fig:RequestFlow} visualizes the output of our request flow construction phase, modeling the bug from the illustrating example (Fig.~\ref{fig:Illustrating Example}). 
The two request flows are shown with distinct background colors. Request spans are represented by colored circles, with arrows indicating the happened-before relationship.
Data spans, which represent operations on shared state stores, appear as character-labeled squares colored by entity identity. In this specific example, the state stores are databases (Order DB and InsidePayment DB), and their names are displayed above the data spans.
The connection between a request and a state access is shown with dashed arrows (solid for writes, hollow for reads). Red highlights and warning symbols mark the potential conflicts identified between the two flows.
To focus the visualization on the race condition, we only show the user requests that initiate the flows and the subsequent request spans that access shared state.

The details captured by \name are critical for analysis. 
For instance, in request flow B, the Cancel Service invokes the Order Service three times (requests \textcircled{4}, \textcircled{5}, and \textcircled{7}). \name distinguishes these requests by their HTTP methods and body content. 
As shown in Fig.~\ref{fig:RequestFlow}, request \textcircled{4} is the only \texttt{GET} request among them. The other two are \texttt{PUT} requests that perform different update operations: 
request \textcircled{5} updates the \texttt{status} field of the order entity to 7 (\texttt{Refunding}), whereas request \textcircled{7} updates it to 4 (\texttt{Cancelled}). Within this single flow, each request is issued only after the previous one has completed, establishing a clear, causally-ordered relationship.

\subsection{Conflicting Request Pair Identification}

After constructing the end-to-end request flows, \name identifies potential concurrency conflicts by analyzing how different requests interact with shared state resources. The core principle is that two requests are potentially in conflict if they access the same shared data entity, and at least one of them performs a write operation.

\subsubsection{Linking Requests to Shared States}
\label{Linking Requests to Database Entities}

The first step is to precisely identify the shared state accessed by each request.
The definition of this state varies across different data stores (e.g., databases, message queues, caches, shared storage) used in microservice systems.
We use the link between request spans and their corresponding data spans, which contain the operational details needed to pinpoint the state.

\begin{figure}[t]
    \centering
    \includegraphics[width=1\linewidth]{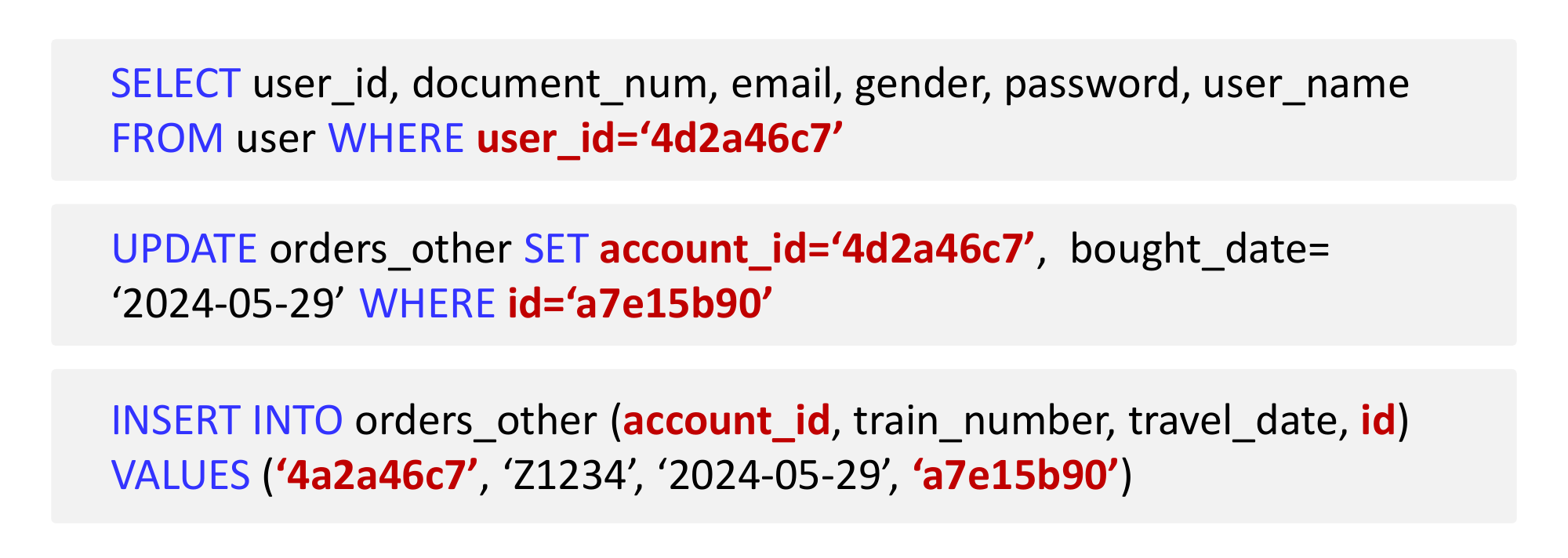}
    \caption{SQL statements executed in TrainTicket during runtime.}
    \label{fig:SQL}
\end{figure}

For traditional relational databases, a state corresponds to a specific row within a table. To pinpoint this row, we identify its primary key from the SQL statements captured in the data span. Since detailed schema information is often unavailable, we employ a heuristic approach that matches field names against common primary key indicators like \texttt{ID/UID/GUID}, \texttt{Key}, or \texttt{Serial}. Developers can also provide custom rules to improve the accuracy of this heuristic.
For \texttt{SELECT}, \texttt{UPDATE}, and \texttt{DELETE} statements, the \texttt{WHERE} clause is parsed to extract the entity-identifying key-value pairs, e.g., \texttt{"accountId=4d2a46c7."}
For \texttt{INSERT} statements, the fields and their corresponding values are extracted directly.
Furthermore, to enable a finer-grained analysis that reduces false positives, we also identify the specific columns involved in the operation. Two operations on the same row only conflict if they access an overlapping set of columns. Therefore, we parse the SQL to find the columns listed in a \texttt{SELECT} clause, the columns being assigned new values in an \texttt{UPDATE} statement's \texttt{SET} clause, or all columns involved in an \texttt{INSERT}.
Fig.~\ref{fig:SQL} illustrates some of the simplified SQL statements executed in the TrainTicket during runtime.
The key-value pairs extracted by \name are highlighted in red. Note that in the \texttt{UPDATE} statement shown in the figure, we extract not only the fields from the \texttt{WHERE} clause (e.g., \texttt{account\_id}) but also other fields (e.g., \texttt{id}). This is because the \texttt{WHERE} clause may use non-primary key conditions to filter the operation targets. Therefore, all ID-related fields in the statement need to be extracted to ensure accurate identification of the database entities affected by the SQL operation.

This principle of identifying a unique state extends to other forms of shared states. For key-value stores like Redis or Memcached, the analysis is more direct, i.e., the entity is the key itself, which we extract from commands like \texttt{GET}, \texttt{SET}, or \texttt{DEL}. For message queues such as Kafka or RabbitMQ, the shared entity is the message stream, identified by the topic or queue name captured from producer or consumer operations. Similarly, for shared object stores like Amazon S3, an entity is a specific object, uniquely identified by its bucket and key, which are recorded from the API call.

After this analysis, each request that interacts with a shared resource is mapped to a precise state, which can be a specific set of database cells, a key-value entry, a message topic, or a storage object. This detailed mapping is the foundation for the subsequent steps of conflict detection.

\subsubsection{Identifying Potential Conflicting Request Pairs}

With the specific state for each request identified, \name systematically filters the vast set of all possible request pairs to find those with a potential for concurrency conflicts. A pair is flagged as a potential conflict only if it satisfies two strict criteria, i.e., the requests must operate on the exact same shared state, and their access patterns must be incompatible (i.e., Read-Write or Write-Write).

The first criterion is to ensure that both requests target the same shared state entity. This requires more than just matching an ID. We enforce two conditions. First, the requests must access the same state store instance, 
verified using information from the data span like the database IP and name. This check is crucial for preventing false positives where different services might use the same ID value (e.g., an \texttt{order\_id}) in their own separate, independent databases. Second, the entity identifiers themselves must match (e.g., the same primary key value, Redis key, or object path). For SQL operations, this means the intersection of their identified primary key sets must be non-empty.
To further refine the accuracy for databases, we apply a column-level check. Even if two requests modify the same row, they are not considered conflicting if they operate on mutually exclusive sets of columns. A conflict is only registered if their sets of accessed columns overlap.
For example, in Fig.~\ref{fig:RequestFlow}, the SQL statements associated with both Request \textcircled{2} and Request \textcircled{4} contain user ID \texttt{4d2a46c7} and order ID \texttt{a7e15b90}, and they access the same column \texttt{status}. Therefore, we consider them to be accessing the same shared state entity.

The second criterion is that the request pair must possess a conflicting access pattern. We classify each data store operation as either a Read or a Write. For databases, \texttt{SELECT} is a Read, while \texttt{INSERT}, \texttt{UPDATE}, and \texttt{DELETE} are Writes. For key-value stores, \texttt{GET} is a Read, while \texttt{SET} and \texttt{DEL} are Writes. Any pair consisting of two Read operations is non-conflicting and is discarded.
Particularly, this classification is based on the actual operation performed at the data store, not the HTTP method of the incoming service request. Relying on HTTP methods is unreliable; for instance, a service handling an HTTP \texttt{GET} request might internally perform a write operation for logging or caching purposes. In Fig.~\ref{fig:RequestFlow}, request \textcircled{6} is an HTTP \texttt{GET} request, but its handling results in an \texttt{INSERT} statement being executed against the database, which is a definitive Write operation. By analyzing the data span, \name correctly identifies this behavior.

\subsubsection{Static Pruning Based on Request Flow}

\name applies two static pruning strategies to refine the set of conflicting request pairs, thereby significantly narrowing the search space for subsequent validation, i.e., \textit{state store instance pruning} and \textit{flow pruning}.

\textbf{State store instance pruning}.
This initial filtering step enforces the fundamental rule that a conflict can only occur if two requests access the exact same state store instance. The identity of an instance is determined using connection metadata captured in the data span, such as the database's IP address and name, the key-value store's endpoint, or the message queue's cluster address.
\name is designed to handle complex real-world topologies. For instance, in systems with master-slave database replication or other clustered data stores, a simple endpoint string match might be insufficient. Therefore, we provide a configurable mechanism that allows developers to define rules for what constitutes the ``same'' logical data store. This ensures, for example, that operations on a primary database and its replica can be correctly identified as interacting with the same shared state.


\textbf{Flow pruning}.
As we depict in Sec.~\ref{sec:End-to-end Request Flows Construction}, after \name constructs and splits request flows, 
all known explicit synchronization relationships are taken into account. 
This includes sequential calls triggered by the same user request, 
causal synchronization relationships via message queue producer-consumer across different user requests, 
and mutual exclusion synchronization via distributed locks.
Requests within the same request flow have a happened-before relationship, and therefore cannot be sent concurrently.
Consequently, even if they access the same database entity, conflicts will not occur. 
Based on this feature, we can further eliminate non-concurrent request pairs.
Since request spans are linked into request flows through the same flow ID, we only need to check whether the requests in a request pair share the same flow ID.

For asynchronous requests initiated by new threads within the service, we have already handled the divergence and marked them as independent request flows, so they will not be eliminated. 
For example, in Fig.~\ref{fig:AsyncFlow}, requests \textcircled{3} and \textcircled{4} are marked as belonging to different request flows, so the request pair (3, 4) will still be retained.

\subsection{Automated Bug Validation}

After identifying suspicious request pairs, \name begins an automated validation process to confirm which pairs manifest as actual concurrency bugs.

\subsubsection{Early Validation for Creation Operations}

The process begins with a fast-pass strategy to identify clear-cut conflicts involving state creation. This check targets pairs where the latter request performs a creation-type operation, such as a database \texttt{INSERT} or a key-value store \texttt{SET} on a new key.

The rationale is that if two requests operate on the same entity identifier and the second one attempts to create it, a conflict is highly probable regardless of the first operation. For example, a \texttt{SELECT} followed by an \texttt{INSERT} on the same entity suggests a query for an object that did not exist moments before its creation, i.e., a logical flaw. Similarly, two consecutive \texttt{INSERT} operations on the same primary key represent a duplicative action that is almost always erroneous, unless the operation is explicitly designed to be idempotent and the system correctly handles the second attempt. This early check efficiently confirms a class of common bugs without requiring full interleaving tests.



\subsubsection{Automated Interleaving Testing}

For the remaining suspicious pairs, \name conducts automated forced interleaving tests to observe if reversing the execution order alters the system's behavior or final state. The testing process for each request pair ($\mathcal{A}$, $\mathcal{B}$) begins by capturing a snapshot of the relevant state store(s) to serve as a clean baseline. The two requests are then replayed in their original order ($\mathcal{A}$ $\rightarrow$ $\mathcal{B}$), and all relevant outputs (e.g., service logs, request responses, and the final state of the affected stores) are collected. Following this forward execution, the state is reverted to the initial snapshot. Next, the requests are replayed in the reverse order ($\mathcal{B}$ $\rightarrow$ $\mathcal{A}$), and the same set of data is collected again. Upon completion, the state is restored one final time, ensuring a clean environment for testing the next suspicious pair.


\subsubsection{The Testing Oracle}

Upon gathering the information from each suspicious request pair through interleaving testing, we submit each pair's data to the oracle. The oracle then executes a three-stage validation process to ascertain whether each pair triggers a concurrency bug. 
This validation process is grounded on the principle of comparing the behavior of request pairs under different execution orders, excluding variables like timestamps, system metrics, or other environment-specific details that do not reflect the impact of order changes.

\begin{itemize}[noitemsep,leftmargin=5.5mm]
    \item \textit{Service-level Validation}: We compare the logs of the involved services under the two interleaving orders of execution. If there is a discrepancy between the two orders, such as a service throwing an exception in one order but not in the other, we can deduce that this request pair triggers a concurrency bug.

    \item \textit{Response-level Validation}: We compare the corresponding responses of the same request across the two orders. If the responses differ, it is evident that the request's outcome is influenced by the execution order, indicating that the pair meets the criteria for concurrency bugs.

    \item \textit{State-level Validation}: We compare the final state snapshots of the involved data stores. Any divergence in the final stored data between the two runs indicates a data inconsistency bug. To reduce false positives in systems with weak consistency models (e.g., a social platform where eventual consistency is acceptable), we provide an optional interface. It allows developers to specify invariants or expected final values for specific data fields, enabling \name to distinguish between benign, business-allowed inconsistencies and true data corruption bugs.
    
\end{itemize}

Through these three stages, conflicting request pairs can be effectively identified. Moreover, the stages are designed to progressively unveil the effects of concurrency bugs from a multi-level perspective. This methodology helps developers understand the impact of concurrency issues, thereby facilitating the remediation of bugs.


%% file: sections/04-implementation.tex
\section{Implementation}
\label{sec:implementation}


We implement \name in Python.
For instrumentation and request flow construction, \name leverages the telemetry capabilities of Apache SkyWalking\cite{apacheskywalking}, a distributed tracing system for cloud service system architectures. Its multi-language instrumentation support (e.g., Java/Python/Go) and extensive library coverage enable \name's multi-language service testing capability. 
Take Java-based services as an example, \name collects request spans by instrumenting HTTP libraries (SpringMVC/RestTemplate) and data spans through JDBC instrumentation. \name can also incorporate trace data collected from middleware components such as cache and message queues into the request flow construction and analysis.
It then automatically extracts request flows from the aggregated trace data, and performs concurrent bug detection.

To support automated interleaving testing, we enhance SkyWalking plugins to log the complete request and response payloads in traces. This includes the full headers and body for HTTP and the method, parameters, and return values for various RPC protocols. This allows \name to precisely replay request contexts during analysis.
Particularly, while implemented with SkyWalking, \name can analyze telemetry data generated by other frameworks (such as OpenTelemetry) with minimal adaptation to the data format.




%% file: sections/05-experiment.tex
\section{Experimental Evaluation}
\label{sec:exp}

In this section, we evaluate the performance of \name for concurrency bug detection.
Particularly, we aim to answer the following two research questions:

\begin{itemize}[noitemsep,leftmargin=5.5mm]
\item \textbf{RQ1}: Can \name identify potential conflict request pairs?
\item \textbf{RQ2}: Can \name detect concurrency bugs in microservice systems?
\end{itemize}

\subsection{Experiment Setting}


\begin{table*}
  \centering
  \caption{Experiments result. ``Early Validation'' shows the number of request pairs that can be confirmed as conflicts early, all considered as positives. ``Auto Test'' shows the number of request pairs that need to be verified through automated interleaving testing. ``FP'' and ``TP'' display the numbers of false positives and true positives, respectively. For true positives, the ``(x+y)'' indicate that x pairs are identified through early validation, while y pairs through automated interleaving testing.}
  \label{tab:experiments result}
  \begin{tabular}{ccccc|ccc|cc|cc}
    \toprule
    \textbf{Case} & \textbf{Benchmark} & \textbf{\makecell[c]{Request \\ Spans}} & \textbf{\makecell[c]{Data \\ Spans}} & \textbf{\makecell[c]{Request \\ Flows}} & \textbf{\makecell[c]{Random \\ Pairs}} & \textbf{\makecell[c]{Instance \\ Pruning}} & \textbf{\makecell[c]{Flow \\Pruning}} & \textbf{\makecell[c]{Early \\Validation}} & \textbf{\makecell[c]{Auto \\Test}} & \textbf{FP} & \textbf{TP}\\
    
    \midrule
    \midrule
    
    1   & TrainTicket       & 63 & 81 & 10 & 78 & -56 & -5 & 6 &  11 & 2     & 11 (6+5)     \\
    2   & TrainTicket       & 56 & 78 & 11 & 76 & -61 & -3 & 5 &  7  & 0     & 9 (5+4)      \\
    3   & TrainTicket       & 55 & 75 & 11 & 68 & -50 & -3 & 6 &  9  & 5     & 9 (6+3)     \\
    4   & TrainTicket       & 69 & 84 & 11 & 9  &  -7 & -1 & 1 &  0  & 0     & 1 (1+0)      \\
    5   & TrainTicket       & 59 & 83 & 13 & 99 & -80 & -3 & 9 &  7  & 6     & 10 (9+1)     \\
    6   & Bank of Anthos    & 10 & 16 & 8 & 28 & -12 & -5 & 5 & 6 & 0 & 7 (5+2)   \\
    7   & Bank of Anthos    & 10 & 16 & 8 & 28 & -12 & -3 & 7 & 6 & 0 & 9 (7+2)   \\
    8   & Bank of Anthos    & 14 & 17 & 9 & 49 & -41 & -1 & 3 & 4 & 3 & 4 (3+1)   \\
    9   & Bank of Anthos    & 15 & 19 & 8 & 36 & -28 & -4 & 4 & 0 & 0 & 4 (4+0)   \\
    10   & SocialNetwork     & 23 & 32 & 9  & 91 & -72 & -3 & 8 &  8  & 3     & 11 (8+3)        \\
    11   & SocialNetwork     & 24 & 34 & 10 & 98 & -76 & -3 & 9 &  10 & 1     & 15 (9+6)          \\
    12   & MediaService      & 20 & 29 &  3 & 64 & -53 & -2 & 4 &  5  & 0     & 6 (4+2)          \\
    13   & MediaService      & 26 & 41 &  5 & 75 & -64 & -2 & 6 &  3  & 2     &  7 (6+1)          \\
    14  & MediaService      & 21 & 32 &  5 & 62 & -51 & -1 & 6 &  4  & 2     &  7 (6+1)          \\
    \bottomrule
  \end{tabular}
\end{table*}

\subsubsection{Microservice Benchmarks}

We select four popular open-source microservice systems for our experiments, including TrainTicket~\cite{DBLP:journals/tse/ZhouPXSJLD21}, Bank of Anthos~\cite{Bank-of-Anthos}, and two microservice applications from DeathStarBench (i.e., Social Network and Media
Service)~\cite{DBLP:conf/asplos/GanZCSRKBHRJHPH19}.
Among them, both the TrainTicket and DeathStarBench have been confirmed to have concurrency issues in previous studies~\cite{DBLP:conf/sosp/LoffP0M023,DBLP:journals/tse/ZhouPXSJLD21} (the bugs in TrainTicket are reproduced from industrial failures).
In addition to the known concurrency bugs, we also inject bugs from previous studies~\cite{DBLP:conf/sigsoft/QiuS0J21} for a more comprehensive evaluation.

\begin{itemize}[noitemsep,leftmargin=5.5mm]
    \item \textbf{TrainTicket:} TrainTicket~\cite{DBLP:journals/tse/ZhouPXSJLD21} is a railway ticketing application based on microservice architecture, which has been widely used in many microservice-related studies~\cite{DBLP:conf/sigsoft/0001ZZIGC22,DBLP:conf/IEEEcloud/ChenJSLZ24}. It consists of 41 services implemented in different programming languages such as Java, Go, Node.js, Python, providing users with functionalities such as ticket query, booking, and payment. Most business functionalities in TrainTicket are implemented using cross-service calls, requiring the collaboration of multiple microservices. The most complex business logic is the ticket booking function, whose call graph involves over ten services.

    \item \textbf{Bank of Anthos:} Bank of Anthos~\cite{Bank-of-Anthos} is a microservice app open-sourced by Google, consisting of five services developed in Java and Python. It simulates core banking operations including account management and payment processing. The benchmark demonstrates typical financial transaction workflows through inter-service communication, with the payment function representing its most complex business logic involving multiple service interactions.

    \item \textbf{DeathStarBench:} DeathStarBench~\cite{DBLP:conf/asplos/GanZCSRKBHRJHPH19} is a popular benchmark suite for microservices, which includes four cloud applications, primarily developed in C++. Our experiments are conducted on the Social Network and Media Service. Social Network is a social application comprising 12 services, using Thrift RPCs to coordinate various services to implement functionalities such as post creation, browsing, and searching.
    Media Service is a review application consisting of 13 services.
    It provides external HTTP endpoints for functionalities such as user registration, review writing, and movie information publishing.
\end{itemize}

\subsubsection{Concurrency Bug Replication}
For TrainTicket~\cite{TrainTicket:v1.0}, we minimize source code modifications based on the original business processes to reproduce the concurrency bugs. For bugs that require specific architectures (e.g., master-slave databases), we also make minimal deployment configuration changes to meet the reproduction requirements.
To collect distributed tracing data, we deploy SkyWalking agents in each service. For the experimental workload scripts, we use the automated request scripts provided by TrainTicket and ensure the workload triggers the concurrency fault-related code.


For the other three benchmarks, we deploy and modify the source code along with database configurations to reproduce the faults. We employ the built-in functional test scripts as workloads since they already provide comprehensive coverage of executable user requests. Similar to TrainTicket, we deploy SkyWalking agents across all services of Bank of Anthos to collect trace data. For the DeathStarBench, since the Social Network and Media services (C++-based systems) are not supported by SkyWalking, we directly utilize their built-in Jaeger trace information for request flow analysis.

\subsection{Experimental Results}
\subsubsection{The Effectiveness of Conflict Request Pair Identification (\textbf{RQ1})}

\begin{figure}
    \centering
    \includegraphics[width=1.0\linewidth]{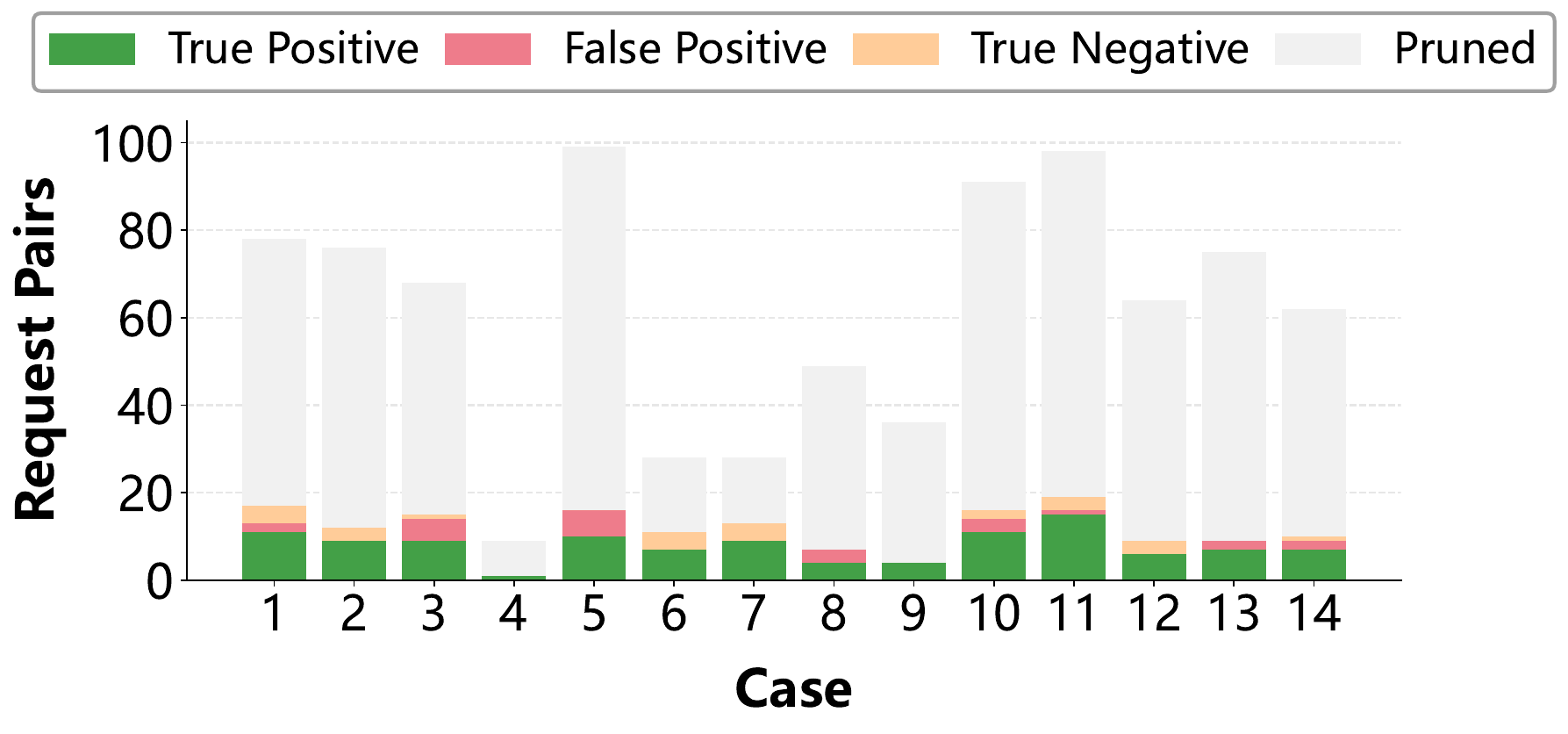}
    \caption{The effectiveness of the pruning strategy. The ``Pruned'' section represents pairs excluded by \name.}
    \label{fig:pruning-rate-figure}
\end{figure}

To answer RQ1, we evaluate how effectively \name prunes the search space for potential conflicts. We establish a baseline using Random Testing, which represents a naive approach where a developer considers all possible request pairs that have at least one write conflict on the same state entity. The total number of such pairs for each case is shown as ``Random Pairs'' in Table~\ref{tab:experiments result}.
Our results demonstrate that \name's two-stage pruning strategy is highly effective. Fig.~\ref{fig:pruning-rate-figure} visually confirms this, showing a stacked bar chart for all 14 test cases. The overwhelming proportion of ``Pruned'' sections (light gray) in most cases illustrates the strategy's ability in narrowing the search space.

Quantitatively, our method reduces the number of candidate pairs requiring validation by an average of 79\% compared to the random baseline. This is achieved through two sequential pruning stages detailed in Table~\ref{tab:experiments result}.
The first stage of pruning leverages state instance information. By enforcing that conflicting requests must target the same state instance, not just a similarly named entity, \name filters out a significant number of false candidates. As shown by the ``Instance Pruning'' column in Table~\ref{tab:experiments result}, this step alone eliminates an average of 73\% of the pairs from the random baseline.
The second stage applies the request flow analysis. Using the happened-before relationships established during flow construction, \name discards pairs that have a sequential relationship within the same flow and thus cannot be concurrent. The ``Flow Pruning'' column in Table~\ref{tab:experiments result} shows this step further removes an additional 6\% of non-concurrent pairs on average.

The small set of pairs remaining after these stages (the green ``True Positive'' and pink ``False Positive'' bars in Fig.~\ref{fig:pruning-rate-figure}) are the high-potential candidates passed to the validation phase. It is worth noting that in Case 4 (TrainTicket), the initial number of ``Random Pairs'' is unusually low (only 9). This is because the specific fault in this case triggers a fatal error, causing the workload to terminate prematurely. The resulting truncated trace contains very few database write operations, naturally limiting the number of potential conflicts.


In summary, by systematically applying constraints based on both state locality and causal dependencies, \name dramatically reduces the search space, making automated validation computationally feasible. Moreover, since it operates at the system level, a single analysis can generate candidate pairs covering multiple distinct concurrency faults, provided the workload has sufficient service coverage.

\subsubsection{The Effectiveness of Concurrency Bug Detection (\textbf{RQ2})}


Our experiments show that \name is effective at detecting concurrency bugs. As shown in the ``TP'' (True Positives) column of Table~\ref{tab:experiments result}, \name successfully identifies all replicated concurrency faults across four benchmarks, with six detected through early validation and five through automated interleaving testing.
Particularly, \name detects a previously unknown bug in the Bank of Anthos benchmark, which is not introduced through our fault injection. We will discuss it in our first case study (Sec.~\ref{sec:case_studies}).

While effective, the framework does produce a small number of false positives (FPs). Our analysis reveals that the primary source of these FPs is a specific and identifiable pattern: benign race conditions involving pure query requests. These are read-only requests intended to retrieve data for frontend displays or to monitor system status, where variations in response data are expected and do not indicate a data corruption bug.
The mechanism causing these FPs lies within our response-level validation oracle. When a pure query request races with a write operation, its response content will naturally differ depending on whether it executes before or after the write. For example:

\begin{itemize}[noitemsep,leftmargin=5.5mm]
    \item In Case 1, the FPs include a request to \texttt{GET /secu-} \texttt{rityservice/securityConfigs/\{accountId\}}. This endpoint queries recent user activity to detect abnormal behavior. While its purpose is for security, its operation is a simple database read. Racing with a write causes its response to change, triggering an FP.
    
    \item In Case 5, the six FPs all involve the request of \texttt{GET /orderOtherService/orderOther/refresh}, which fetches all of a user's order data. When this races with a request that modifies an order, the list of returned orders changes, which our oracle flags as an inconsistency.
\end{itemize}

This behavior is a known limitation of automated validation that relies on stateless response comparison. To address this and empower developers to refine the analysis, \name includes a practical whitelist mechanism. Testers can use this feature to provide domain-specific knowledge by marking specific read-only endpoints where response variations are benign and expected. By exempting these configured interfaces from response-level validation, the number of false positives can be significantly reduced with minimal manual effort.

\subsection{Case Studies}
\label{sec:case_studies}

To illustrate the practical application and effectiveness of \name, we present three concrete case studies.

\begin{itemize}[noitemsep,leftmargin=5.5mm]
\item \textbf{Case 1:}
This case study highlights a classic check-then-act race condition within the Bank of Anthos benchmark. The vulnerability arises when two valid, independent transfer requests are processed concurrently by the LedgerWriter service.
The service's logic first performs a balance check (the ``check'') and then executes the transfer (the ``act''). However, without proper locking or atomic operations, a critical window exists between these two steps. This allows both requests to read the initial balance and pass the funds validation before either can update the account. Consequently, both transfers are committed, leading to an unauthorized overdraft. This bug bypasses the existing idempotency logic, which is not designed to enforce atomicity for concurrent, non-duplicate operations.

\item \textbf{Case 2:} This case demonstrates the detection of the classic inter-service race condition between payment and cancellation flows, as detailed in Sec.~\ref{Illustrating Example}.
\name identifies the conflicting request pairs (highlighted in Fig.~\ref{fig:RequestFlow}) and subjects them to automated interleaving tests. The tests reveal a critical behavioral inconsistency: the service's HTTP response for the cancellation operation differs depending on the execution order. This allows the response-level validation mechanism to confirm the bug, correctly reporting both pairs as true positives.


\item \textbf{Case 3:} This case demonstrates a classic read-modify-write atomicity violation. In the replicated TrainTicket scenario, two requests initiated via the Preserve Service attempt to reserve the last available seat from the Seat Service concurrently. A race condition occurs when the first request reads the seat availability (count = 1) but is delayed before it can write the update (decrementing the count to 0). During this critical window, the second request also reads the seat count as 1 and proceeds. Consequently, both purchase requests succeed, causing the seat to be oversold and violating data integrity.


\end{itemize}

\section{Threats to Validity}
\label{sec:validity}


\textbf{External Validity.} The generalizability of our findings is constrained by our evaluation's reliance on a specific set of open-source microservice benchmarks. These systems may not capture the full diversity of architectures, workloads, or data store technologies used in all industrial applications. To mitigate this, we select benchmarks that are complex, widely used in both academic and industrial research, and considered representative of common microservice patterns. Furthermore, the bugs we reproduce reflect prevalent and challenging concurrency issues, suggesting that \name's effectiveness is likely applicable to a broader range of real-world systems.

\textbf{Construct Validity.} The primary threat to construct validity concerns the accuracy with which \name identifies shared entities and confirms bugs.

First, identifying the specific entity being accessed relies on a heuristic to parse primary key values from SQL statements. This approach can be imprecise for highly complex queries or those that filter on non-primary-key attributes (e.g., \texttt{SELECT * FROM orders WHERE order\_time < '2024-06-01';}). This may cause \name to miss certain conflicts, leading to potential false negatives. We acknowledge this as a fundamental challenge of SQL analysis. Therefore, our design allows developers to refine heuristics when ambiguities arise.

Second, \name's black-box nature presents a risk of false positives. It detects data races at the data store level but cannot inspect an application's internal source code. A service might implement its own synchronization logic (e.g., optimistic locking using version numbers or Compare-and-Swap operations) that correctly handles a data race. Because this code-level logic is invisible to our instrumentation, \name may flag such a benign race as a bug. While our multi-level validation oracle helps filter many cases, this remains a limitation inherent to any black-box testing approach.

%% file: sections/06-related_work.tex
\section{Related Work}
\label{sec:related_work}


\textbf{Static Analysis. } Static analysis emerges as a proactive technique for identifying potential concurrency issues without the necessity of code execution. For example, to identify deadlocks that may arise from the integration of test control mechanisms, Chen et al.~\cite{DBLP:conf/edo/Chen00} proposed an approach that leverages static analysis to construct test constraints, with a particular focus on synchronization events. Lu et al.~\cite{DBLP:journals/tse/LuLLLFX22} introduced a tool that combines static analysis, log mining~\cite{DBLP:journals/csur/HeHCYSL21}, and log enhancement techniques to efficiently detect concurrency bugs in distributed systems. By converting concurrency bug detection into a source-sink reachability problem, Cai et al.~\cite{DBLP:conf/pldi/CaiYZ21} presented a value-flow analysis framework for statically finding diversified inter-thread value-flow bugs. These methods play a crucial role in the early detection of problems during the development phase, thus preemptively addressing issues that could lead to runtime errors. However, static methods do not always excel at detecting concurrent bugs, given the inherent complexity of the runtime environment in distributed systems.

\textbf{Dynamic Analysis.}
Dynamic analysis plays a pivotal role in the detection of concurrency issues by executing the system and monitoring its runtime behaviors.
It provides a real-time perspective on how concurrent operations interact within a system. 
For example, built with a set of happens-before rules that model various communication and concurrency mechanisms, DCatch~\cite{DBLP:conf/asplos/LiuLLLLGT17} predicts distributed concurrency bugs by analyzing correct execution of distributed systems. Similar to our work, ReqRacer~\cite{DBLP:conf/sigsoft/QiuS0J21} employs the construction of a dependency graph to identify potential conflict-prone request pairs for interleaving testing. Yuan et al.~\cite{DBLP:conf/asplos/YuanY20} used partial order sampling and conflict analysis to efficiently test Erlang-based systems. Focusing on scalability, FlyMC~\cite{DBLP:conf/eurosys/LukmanKSSKSPTYL19} presents a tool for scalable testing of complex interleavings in distributed systems to detect concurrency bugs. Meanwhile, Wu et al.~\cite{DBLP:conf/ica3pp/WuLW15} targeted efficiency by identifying repeated interleavings, which is a pattern often found in concurrency bug detectors.
However, these approaches are tailored for traditional distributed systems and do not address the challenges unique to microservice architecture, which often employs a ``database-per-service'' design pattern.


\textbf{Machine Learning and AI.}
Recent advancements in AI and machine learning have been applied to concurrency testing to improve the efficiency and effectiveness of detecting issues. Zhang et al.~\cite{DBLP:conf/sigsoft/ZhangWLQRZ14} presented a system that uses AI to tolerate concurrency bugs, which can be particularly useful in distributed environments where bugs are hard to eradicate. Mukherjee et al.~\cite{DBLP:journals/pacmpl/MukherjeeDBL20} introduced a framework called QL that uses reinforcement learning to explore the space of possible interleavings in concurrent programs, adapting to the program under test. While seemingly general and powerful, Machine Learning and AI techniques often demand substantial computational resources for training. Additionally, they introduce a level of uncertainty due to their nature as black box models.


%% file: sections/07-conclusion.tex
\section{Conclusion}
\label{sec:conclusion}

In this work, we introduce \name, an non-intrusive and automated framework for detecting concurrency bugs in microservice systems.
\name first leverages dynamic instrumentation of widely-used libraries, effectively capturing detailed trace data without modifications to the application code.
These data enable the analysis of happened-before relationships and resource access patterns among common operations in service systems.
\name then identifies suspicious request pairs to check the correctness of their concurrency behaviors.
In this process, \name applies static pruning strategies to improve the search efficiency. 
Finally, \name validates the concurrency bugs through a three-stage process.
Overall, \name can effectively address the challenges of complex dependencies and the need for accurate, automated bug validation.
Experiments on four widely-used open-source microservice benchmarks with replicated industrial bugs demonstrate \name's ability to accurately and efficiently detect concurrency bugs.